\documentclass[
twocolumn, 
 amsmath,amssymb,
 aps,
prl,
superscriptaddress
]{revtex4-2}

\usepackage{graphicx}
\usepackage{dcolumn}
\usepackage{bm}
\usepackage[colorlinks=true, allcolors=blue]{hyperref}
\usepackage{braket}
\usepackage{xcolor}
\usepackage{bbding}
\usepackage{natbib}
\usepackage{gensymb}
\usepackage{balance}

\begin{document}
\preprint{APS/123-QED}

\title{Topological switching in bilayer magnons via electrical control}

\author{Xueqing Wan}
\email{These authors contributed equally to this work.}
\affiliation{%
Ministry of Education Key Laboratory for Nonequilibrium Synthesis and Modulation of Condensed Matter, Shaanxi Province Key Laboratory of Advanced Functional Materials and Mesoscopic Physics, School of Physics, Xi'an Jiaotong University, Xi'an 710049, China 
}%
\author{Quanchao Du}
\email{These authors contributed equally to this work.}
\affiliation{%
Ministry of Education Key Laboratory for Nonequilibrium Synthesis and Modulation of Condensed Matter, Shaanxi Province Key Laboratory of Advanced Functional Materials and Mesoscopic Physics, School of Physics, Xi'an Jiaotong University, Xi'an 710049, China 
}
\affiliation{%
 Division of Physics and Applied Physics, School of Physics and Mathematical Sciences, Nanyang Technological University, Singapore 637371, Singapore 
}%

\author{Jinlian Lu}
\email{These authors contributed equally to this work.}
\affiliation{%
Department of Physics, Yancheng Institute of Technology, Yancheng, Jiangsu 224051, China 
}%
\author{Zhenlong Zhang}
\affiliation{%
Ministry of Education Key Laboratory for Nonequilibrium Synthesis and Modulation of Condensed Matter, Shaanxi Province Key Laboratory of Advanced Functional Materials and Mesoscopic Physics, School of Physics, Xi'an Jiaotong University, Xi'an 710049, China 
}%
\author{Jinyang Ni}
\email{jyni18@fudan.edu.cn}
\affiliation{%
 Ministry of Education Key Laboratory for Nonequilibrium Synthesis and Modulation of Condensed Matter, Shaanxi Province Key Laboratory of Advanced Functional Materials and Mesoscopic Physics, School of Physics, Xi'an Jiaotong University, Xi'an 710049, China 
}%
\affiliation{ State Key Laboratory of Surface Physics and Department of Physics, Fudan University, Shanghai 200433, China}%

\author{\linebreak Lei Zhang}
\email{zhangleio@xjtu.edu.cn}
\affiliation{%
Ministry of Education Key Laboratory for Nonequilibrium Synthesis and Modulation of Condensed Matter, Shaanxi Province Key Laboratory of Advanced Functional Materials and Mesoscopic Physics, School of Physics, Xi'an Jiaotong University, Xi'an 710049, China 
}%
\author{Zhijun Jiang}
\email{zjjiang@xjtu.edu.cn}
\affiliation{%
 Ministry of Education Key Laboratory for Nonequilibrium Synthesis and Modulation of Condensed Matter, Shaanxi Province Key Laboratory of Advanced Functional Materials and Mesoscopic Physics, School of Physics, Xi'an Jiaotong University, Xi'an 710049, China }%
\affiliation{ State Key Laboratory of Surface Physics and Department of Physics, Fudan University, Shanghai 200433, China}

\author{Laurent Bellaiche}
\affiliation{%
Smart Ferroic Materials Center, Physics Department and Institute for Nanoscience and Engineering, University of Arkansas, Fayetteville, Arkansas 72701, USA
}%
\affiliation{%
Department of Materials Science and Engineering, Tel Aviv University, Ramat Aviv, Tel Aviv 6997801, Israel
}%




\begin{abstract} 
Topological magnons, quantized spin waves featuring 
nontrivial boundary modes, present a promising route toward lossless information processing. Realizing practical devices typically requires magnons excited in a controlled manner to enable precise manipulation of their topological phases and transport behaviors. However, their inherent charge neutrality and a high frequency nature pose a significant challenge for nonvolatile control, especially via electric means. Herein, we propose a general strategy for electrical control of topological magnons in bilayer ferromagnetic insulators. With strong spin-layer coupling, an applied vertical electric field induces an interlayer potential imbalance that modifies intralayer Heisenberg exchanges between adjacent layers. This electric-field-driven modulation competes with the bilayer's intrinsic Dzyaloshinskii-Moriya interaction, enabling the accurate tuning of the band topology and nonreciprocal dynamics of magnons. More importantly, such an electric control mechanism exhibits strong coupling with external magnetic fields, unveiling new perspectives on magnetoelectric coupling in charge-neutral quasiparticles.
\end{abstract}
\maketitle

\textit{Introduction.} 
Over the past decades, the field of solid physics has witnessed the emergence of geometry and topology as fundamental aspects of quantum states in Bloch band theory, culminating in the discovery of Chern insulators that exhibit the quantum anomalous Hall effect\,\cite{thouless1982quantized,haldane1988model,kane2005z,bernevig2006quantum,hasan2010colloquium, yu2010quantized,chang2013experimental, bansil2016colloquium, senthil2015symmetry, vsmejkal2018topological}. These topological concepts extend beyond electrons to bosonic quasiparticles, including phonons\,\cite{xu2024catalog,stenull2019signatures,zhang2010topological} and magnons\,\cite{mcclarty2022topological, chen_PRX_2018_8, mook_PRX_2021_11}, the latter being the quantized spin-wave excitations in ordered magnets. Unlike phonons, the time-reversal symmetry of magnons can be broken by asymmetric or bond-dependent spin exchanges, resulting in the magnonic Chern insulator\,\cite{ruckriegel2018bulk, plekhanov2017floquet, zhu2021topological, mcclarty2022topological}. The magnonic Chern insulator was experimentally observed in layered honeycomb magnets, where the second nearest neighbor\,(2NN) Dzyaloshinskii-Moriya interaction\,(DMI) opens a topological nontrivial gap at the Dirac point, protected by chiral edge states\,\cite{chen_PRX_2018_8,zhu2021topological,zhang_thermal_Physreport_2024_1070}. The 2NN DMI plays a role analogous to complex hopping in the Haldane model, serving as the key mechanism that drives magnon anomalous Hall effect\,\cite{onose2010observation, katsura_prl_2010_104,zhang_PRL_2021_127, mook2014magnon}. Combining charge neutrality and terahertz-scale dynamics, topological magnons present a compelling platform for next-generation devices that can circumvent the energy dissipation limits of conventional electronics\,\cite{chumak2015magnon,lenk2011building,prabhakar2009spin,pirro2021advances}.
\begin{figure*}
    	\centering
    	\includegraphics[scale=0.7]{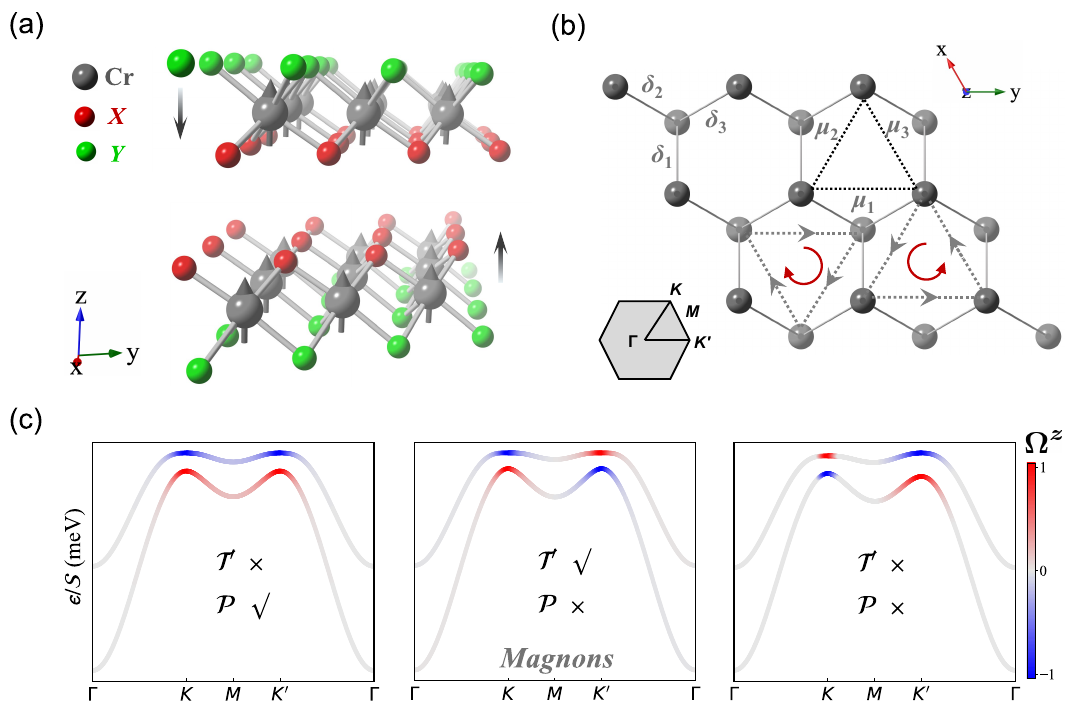}
    	\caption{(a) Side view of AB-stacking Janus bilayer $\mbox{Cr}XY$, where upper (down) vector represents the layer polarization in the first (second) layer. (b) Top view of the magnetic lattice in this bilayer, showing its analogy to a honeycomb lattice. The clockwise and counterclockwise arrows indicate the chirality of the DMI between different sublattices. (c) Magnon bands with distinct symmetry regimes: ${\cal T}^{\prime}$ breaking, ${\cal P}$ breaking, and joint ${\cal PT}^{\prime}$ symmetry breaking.}
    \label{fig1}
\end{figure*}

Ideally, for practical applications, magnons need to be excited in a controlled manner, precisely manipulated and detected, preferably through nonvolatile means\,\cite{chumak2014magnon, chumak2015magnon,arakawa2019control,shen2023electrical,venema2016quasiparticle}. As low-energy spin excitations in ordered magnets, magnons are particularly susceptible to external magnetic fields\,\cite{kuzmenko2018switching,silva2020magnon,li2016weyl, mook_PRX_2021_11}. However, uniform magnetic fields merely induce a global Zeeman shift in the magnon bands, without altering their topological phase or the associated transport properties\,\cite{go2024magnon,ni2025magnon,mukherjee_PRB_2023_107,kondo_prreserach_2022_4}. More critically, from the energy-efficiency perspective, magnetic field control is inherently limited, as 1 Tesla\,(T) yields only around 0.1\,$\mbox{meV}$ of energy shift, which is insufficient for device-level applications. Strain engineering offers an alternative by modulating spin exchange interactions to tune the magnon band topology\,\cite{shrestha2025tunable,Sadovnikov2018,Li2021Magnonic,Vidal-Silva2022,Soenen2023,Zhuo2021}. Nevertheless, it is often constrained by mechanical instability and integration challenges. On the other hand, although the electric field offers greater stability and efficiency\,\cite{Li2021Magnonic,Vidal-Silva2022,Soenen2023,Zhuo2021}, the charge neutrality of magnons poses a significant challenge for direct manipulation. These challenges underscore the urgent need for a new control paradigm that combines high efficiency with precise topological tunability\,\cite{chumak2014magnon, chumak2015magnon, bader2010spintronics}.

In this work, we address this challenge by proposing a novel type of spin-layer coupling to realize electrically controlled topological magnons. Combining density functional theory\,(DFT) calculations and effective model analysis, we demonstrate that an applied vertical electric field ${\cal E}_{z}$ can effectively modulate the layer polarization in bilayer magnets with strong spin-layer coupling, enabling precise control over the intralayer Heisenberg spin exchanges. This electric-field-driven modulation competes with their intrinsic DMI, driving the magnonic phase from the Chern insulator to the trivial insulator. In addition, applying ${\cal E}_{z}$ significantly reduces the required threshold of external magnetic field for topological band manipulation down to the order of 10\,$\mbox{mT}$, offering a novel pathway to achieving magnetoelectric coupling in topological magnons.   

\begin{figure*}
    	\centering
    	\includegraphics[scale=0.725]{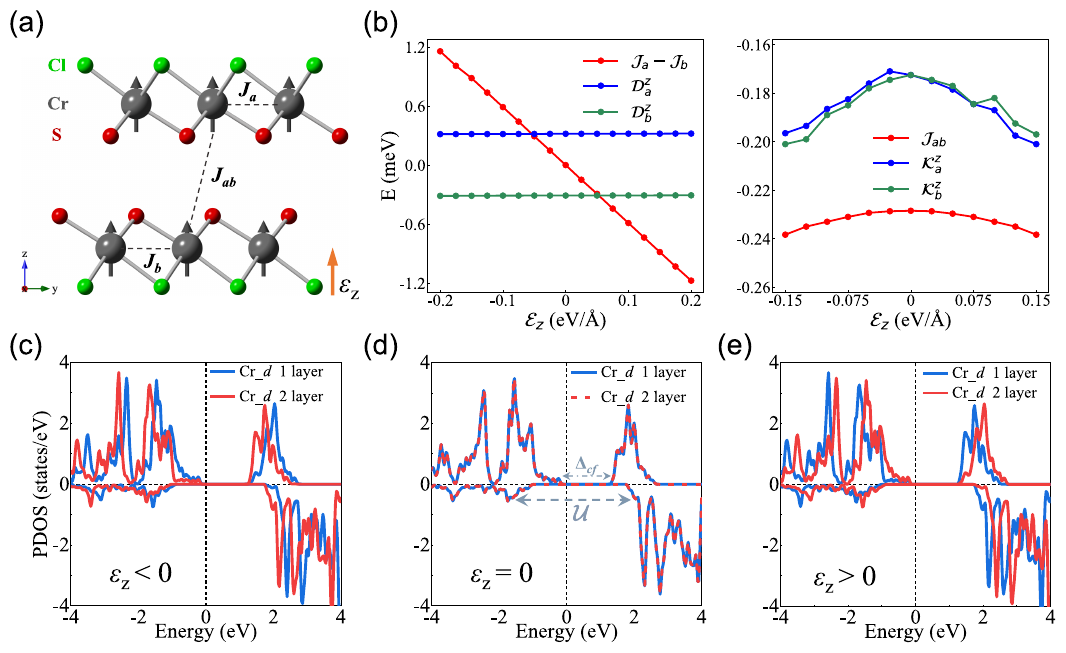}
    	\caption{{The spin exchanges of bilayer $\mbox{CrSCl}$ under ${\cal E}_{z}$.} (a) The illustration of bilayer $\mbox{Cr}XY$ and corresponding spin exchange paths. (b) The dependence of spin exchange parameters on ${\cal E}_{z}$, with left panel illustrating to the relative Heisenberg spin exchanges \,(${\cal J}_{a}-{\cal J}_{b}$), ${\cal D}^{z}_{a}$ and ${\cal D}^{z}_{b}$, while right panel presents the interlayer spin exchange ${\cal J}_{ab}$ and SIA. The projected-density-of-states (PDOS) of Cr$^{3+}$ in the bilayer CrSCl with (c) ${\cal E}_{z}$\,$=$\,$-0.2$, (d) ${\cal E}_{z} = 0$, (e) ${\cal E}_{z}$\,$=$\,$0.2$\, $\text{eV}/\text{\AA}$, respectively.}
    \label{fig2}
\end{figure*}

\textit{Symmetry analysis of magnons.} To investigate the symmetry of magnons, it is essential to
consider both the spin Hamiltonian and its classical magnetic ground state\,\cite{mook_PRX_2021_11, ni2025magnon}. In centrosymmetric honeycomb ferromagnets, for instance, the Dirac points of magnon bands are stable, even though their ferromagnetism violates actual $\cal {T}$ symmetry (or spin flip operation). This distinction from electrons arises primarily from the time-reversal symmetry of magnons defined as $C_{2\boldsymbol{n}}{\cal T}$, where $C_{2\boldsymbol{n}}$ represents the $\pi$ spin rotations about a $\boldsymbol{n}$ axis normal to the magnetization. Here, we refer to $C_{2\boldsymbol{n}}{\cal T}$ as the effective time-reversal symmetry ${\cal T}^{\prime}$. As illustrated in Fig.\,\ref{fig1}(b), ${\cal T}^{\prime}$ can be broken by 2NN DMI in honeycomb ferromagnets, which opens a topologically nontrivial gap at the Dirac points. Conversely, as illustrated in Fig.\,\ref{fig1}(c), the breaking of inversion symmetry ${\cal P}$\, (typically through sublattice asymmetry) induces a trivial gap. The interplay between the ${\cal P}$ and ${\cal T}^{\prime}$ symmetry breaking governs their topological phase\,\cite{haldane1988model,kim2022topological}.

\textit{Spin-layer coupling.} As discussed above, controlling the ${\cal PT^{\prime}}$ symmetry is crucial for manipulating magnonic topological phases. Since ${\cal T}^{\prime}$ symmetry is
linked to the intrinsic spin-orbital coupling\,(SOC), its direct manipulation is inherently challenging. In contrast, recent studies demonstrate that in multilayer magnets with strong spin-layer coupling, the interplay between layer polarization and electric fields provides a novel pathway for nonvolatile manipulation of magnets\,\cite{Zhang2024Predictable, Tian2025, shen2024}. Specifically, as shown in Fig.\,\ref{fig1}(a), in centrosymmetric bilayer magnets with opposing layer polarization, ${\cal E}_{z}$ can significantly modulate the spin polarization in each layer, enabling the coupling between spin and layer-index\,\cite{fiebig2005revival,ni2025magnon}. Such interlayer potential imbalance induced by ${\cal E}_{z}$ further induces asymmetric variations of Heisenberg spin exchange in each layer, which breaks ${\cal P}$, enabling precise control of magnons. 

\begin{table}[b]
\caption{\label{tab_1}Calculated DMI and SIA of Janus bilayer $\mbox{Cr}{XY}$.}
\begin{ruledtabular}
\begin{tabular}{cccccr}
${\cal D}/{\cal K}(\mbox{meV})$  & \multicolumn{1}{c}{${\cal D}^{z}_{a/b}$}  & \multicolumn{1}{c}{${\cal D}^{y}_{a/b}$}  & \multicolumn{1}{c}{${\cal D}^{x}_{a/b}$} & \multicolumn{1}{c}{${\cal K}^{z}_{a/b}$}  \tabularnewline
\hline 
$\mbox{CrSCl}$   & $\pm$0.35 & $\pm$0.03 & $\pm$0.01 &$-$0.17 \tabularnewline
$\mbox{CrSBr}$   & $\pm$0.17  & $\mp$0.15 & $\mp$0.26 &$-$0.23  \tabularnewline
$\mbox{CrSeCl}$  & $\mp$0.14  & $\pm$0.04 & $\pm$0.15  & 0.21  \tabularnewline
$\mbox{CrSeBr}$  & $\pm$0.07  &$\pm$0.15 & $\pm$0.49 &0.11  \tabularnewline
\end{tabular}
\end{ruledtabular}
\end{table}

Following the above symmetry requirement, we identify the Janus bilayer family $\mbox{Cr}XY$\,($X$ = $\mbox{S}$, $\mbox{Se}$; $Y$ = $\mbox{Cl}$, $\mbox{Br}$) as material candidates by employing DFT calculations, as described in Supplemental Material\,\cite{Supplemental_Materials} (see also Refs.\,\cite{winker, PhysRevB.47.14932, huang2018toward, blochl1994projector, perdew1996generalized, phonopy-phono3py-JPCM, phonopy-phono3py-JPSJ, liechtenstein1995density, kresse1996efficiency, xiang2011predicting, grimme2010consistent, xiang2013magnetic, chang1996berry, xiao2010berry, ni2025nonvolatile, bhowal2021orbital, go2024magnon, onose2010observation, katsura_prl_2010_104} therein). The Janus monolayer $\mbox{Cr}XY$ is derived from the $1\mbox{T}$-$\mbox{CrTe}_{2}$ monolayer by replacing one of the chalcogen layers with halides. This substitution drives the monolayer into an insulating state characterized by $\mbox{Cr}^{3+}$ ions with $S$\,$=$\,$\frac{3}{2}$\,\cite{zhang2021room, meng2021anomalous, xian2022spin, hou2022multifunctional, xiao2020two, ni2025nonvolatile}. The phonon calculations demonstrate that bilayer $\mbox{Cr}XY$ can be stabilized in an AB-stacking configuration while maintaining ${\cal P}$ symmetry, as shown in Fig.\,\ref{fig1}(a). Clearly, the magnetic lattice of this bilayer can be mapped onto a monolayer honeycomb ferromagnetic lattice, where the NN interlayer and intralayer spin exchanges correspond to the NN and 2NN intralayer spin exchanges in the honeycomb lattice, respectively. Our DFT calculations reveal that both the NN interlayer (${\cal J}_{ab}$) and intralayer spin exchanges (${\cal J}_{a/b}$) between $\mbox{Cr}^{3+}$ ions exhibit ferromagnetic\,(FM) coupling\,(see Fig.\,\ref{fig2} and Fig.\,S4 \cite{Supplemental_Materials}), confirming the FM ground state. As summarized in Tab.\,\ref{tab_1}, bilayer $\mbox{CrS}Y$ prefers easy axis magnetic anisotropy, while
$\mbox{CrSe}Y$ adopts an easy plane configuration\,\cite{Supplemental_Materials}. Notably, we find that the sizable NN intralayer DMI with opposite DMI vectors in adjacent layers, whose characteristics mirror 2NN DMI of monolayer honeycomb lattice\,\cite{haldane1988model, chen_PRX_2018_8}.

The applied ${\cal E}_{z}$ naturally breaks the ${\cal P}$ of this bilayer, leading to the interlayer imbalance between the adjacent layers. This imbalance is reflected in the remarkable changes in the ${\cal J}_{a}$ and ${\cal J}_{b}$, as illustrated in Fig.\,\ref{fig2}(b), where the calculated results of bilayer $\mbox{CrSCl}$ demonstrate
${\cal J}_{a}$\,$-$\,${\cal J}_{b}$\,$=$\,$-0.6\,\mbox{meV}$ when ${\cal E}_{z}$\,$=$\,$0.1$\,$\mbox{eV/\AA}$. As
the strength of ${\cal E}_{z}$ increases, ${\cal J}_{a}$\,$-$\,${\cal J}_{b}$
exhibits a linear dependence on ${\cal E}_{z}$, described by ${\cal J}_{a}$\,$-$\,${\cal J}_{b}$\,$=$\,$-$6\,${\cal E}_{z}$, highlighting that ${\cal E}_{z}$ can reverse the sign of ${\cal J}_{a}$\,$-$\,${\cal J}_{b}$. To elucidate this ${\cal E}_{z}$ modulation, we derive the formula of spin exchange energy\,($E_{ex}$) between $\mbox{Cr}^{3+}$ ions based on the two-orbitals model. Within second perturbation theory\,\cite{winker, huang2018toward}, $E_{ex}$ can be expressed as
\begin{equation}\label{eq1}
     E_{ex} = -\frac{2 t^{2}}{\Delta_{cf}} + 2t^{2}\left(\frac{1}{\Delta_{cf}+{\cal U}}+\frac{1}{\cal U}\right),
\end{equation}
where $t$ is the hopping term, ${\Delta_{cf}}$ represents the crystal field splitting, and ${\cal U}$ refers to the spin splitting between spin up and down states. Generally, in $3d$ magnetic systems with high spin states, ${\cal U}$\,$\gg$\,$\Delta_{cf}$; thus, we can obtain the relation between $E_{ex}$ and $\Delta_{cf}$ for fixed ${\cal U}$ and $t$, as shown in Fig.\,S6\,\cite{Supplemental_Materials}, demonstrating that $E_{ex}$\,$\propto$\, $1/\Delta_{cf}$. Given that NN intralayer ${\cal J}$ depends on $E_{ex}$, the above alternating behavior between ${\cal J}_{a}$ and ${\cal J}_{b}$ induced by ${\cal E}_{z}$ can be attributed to changes in $\Delta_{cf}$ in $\mbox{Cr}^{3+}$ ions within each layer. Fig.\,\ref{fig2}(c)-(e) show the evolution of the PDOS of bilayer $\mbox{CrSCl}$ under ${\cal E}_{z}$. In the absence of ${\cal E}_{z}$, the $\Delta_{cf}$ of $\mbox{Cr}^{3+}$ ions in both layers is identical, enforcing ${\cal J}_{a}$\,$=$\,${\cal J}_{b}$. However, when ${\cal E}_{z}$\,$<$\,$0$, the first layer exhibits a larger $\Delta_{cf}$ than the second layer, leading to $|{\cal J}_{a}|$\,$<$\,$|{\cal J}_{b}|$. Conversely, a positive ${\cal E}_{z}$ reverses this trend, hence $|{\cal J}_{a}|$\,$>$\,$|{\cal J}_{b}|$. Unlike ${\cal J}_{a/b}$, as shown in Fig.\,\ref{fig2}(b), ${\cal D}_{z}$ remains nearly unaffected by ${\cal E}_{z}$, as do the single ion anisotropy (SIA) and interlayer exchange coupling\,(${\cal J}_{ab}$). The distinct responses to the electric field, especially between ${\cal J}_{a/b}$ and ${\cal D}_{z}$, give rise to electrically tunable magnons in this bilayer.

\textit{Easy axis bilayer.} Next, we investigate the influence of ${\cal E}_{z}$ on magnons in bilayer $\mbox{CrS}Y$, where the minimal spin Hamiltonian can be expressed as
\begin{equation}
\begin{split}
\label{eq2}
{\cal \hat{H}} = & {\cal J}_{ab}\sum_{\langle i,j\rangle}{\cal S}_{i}\cdot{\cal S}_{j} + \sum_{\langle i,j \rangle,\alpha} {\cal J}_{\alpha}\left({\cal S}_{i,\alpha} \cdot {\cal S}_{j,\alpha}\right) + \\
& \sum_{\langle i,j \rangle,\alpha} {\cal D}^{z}_{\alpha}\cdot\left({\cal S}_{i,\alpha} \times {\cal S}_{j,\alpha}\right) + \sum_{ i,\alpha}{\cal K}^{z}_{\alpha}\left({\cal S}^{z}_{i,\alpha}\right)^{2}.
\end{split}
\end{equation}
Here, the first term represents the NN interlayer spin exchanges, and the second to fourth terms refer to the NN intralayer spin exchanges, out-of-plane DMI, and SIA within $\alpha$-index-layers\,($\alpha$\,$=$\,$a,b$), respectively. Given the weak response of ${\cal J}_{ab}$, SIA, and DMI to ${\cal E}_{z}$, we can assume that these parameters remain invariant under the applied electric field in the following discussion. Upon Holstein-Primakoff\,(HP) transformation and Fourier
transformation\,\cite{holstein1940field}, one can obtain the linear magnon Hamiltonian of Eq.\,(\ref{eq2}) in $k$-space, ${\cal {\hat{H}}}=\sum_{k}\psi_{k}^{\dagger}{\cal \hat{H}}_{k}\psi_{k}$, where $\psi_{k}^{\dagger}\equiv(\hat{a}_{k}^{\dagger},\hat{b}_{k}^{\dagger})$
and ${\cal \hat{H}}_{k}$\,$=$\,$h_{0}\sigma_{0}$\,$+$\,$\boldsymbol{h}$\,$\cdot$\,$\boldsymbol{\sigma}$.
Here, $h_{0}$\,$=$\,$3{\cal J}_{ab}$\,$+$\,$\frac{1}{2}({\cal J}_{a}$\,$+$\,${\cal J}_{b})f_{k}$\,$-$\,$2{\cal K}_{z}$, $h_{x}$\,$=$\,${\cal J}_{ab}\mbox{Re}\gamma_{k}$, $h_{y}$\,$=$\,${\cal J}_{ab}\mbox{Im}\gamma_{k}$
and $h_{z}$\,$=$\,$\frac{1}{2}({\cal J}_{a}$\,$-$\,${\cal J}_{b})f_{k}$\,$+$\,${\cal D}_{z}d_{k}$. Note that  $\gamma_{k}$\,$=$\,$\sum_{\delta}e^{-i\textbf{\textit{k}}\cdot\boldsymbol{\delta_{i}}}$, $f_{k}$\,$=$\,$\sum_{i\in odd}2\mbox{cos}(\textbf{\textit{k}}\cdot\boldsymbol{\mu}_{i})$\,$-$\,$6$ and $d_{k}$\,$=$\,$\sum_{i\in odd}2\mbox{sin}(\textbf{\textit{k}}\cdot\boldsymbol{\mu}_{i})$, where $\boldsymbol{\delta}_{i}$ and $\boldsymbol{\mu}_{i}$ represent the
linking vectors of honeycomb lattice as shown in Fig.\,\ref{fig1}(b).
\begin{figure}
\centering
\includegraphics[scale=0.45]{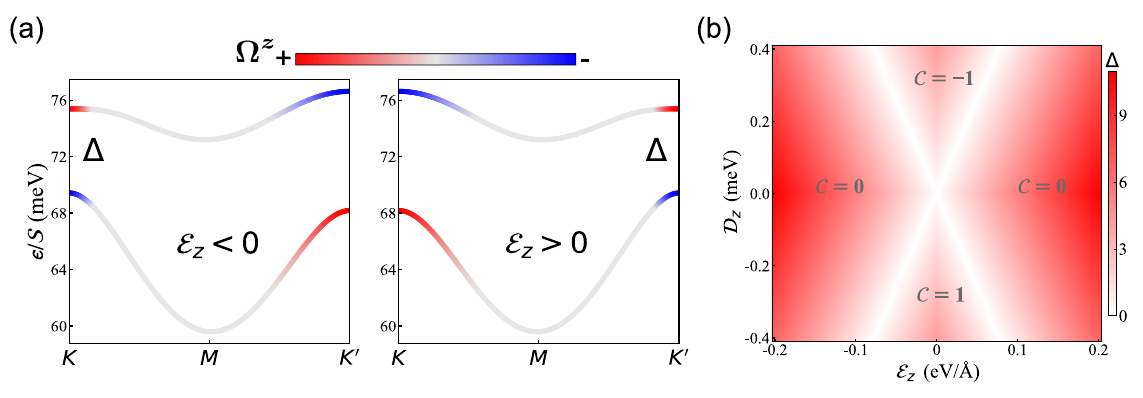}
    	\caption{The magnon nonreciprocity and phase diagram. (a) The magnon bands of Janus bilayer $\mbox{CrSCl}$ with ${\cal E}_{z}$\,$<$\,$0$ and ${\cal E}_{z}$\,$>$\,$0$. (b) The phase diagram of magnons in a Janus bilayer with easy-axis anisotropy.}
\label{fig3}
\end{figure}

For simplicity, the 
$\boldsymbol{h}(\boldsymbol{k})=\left(h_{x},h_{y},h_{z}\right)$ can be parameterized as $|h|\left(\mbox{sin}\theta\mbox{cos}\phi,\mbox{sin}\theta\mbox{sin}\phi,\mbox{cos}\theta\right)$.
The corresponding eigenvalues and eigenvectors are thus given by 
\begin{equation}\label{eq3}
{\epsilon}_{\pm}/S=h_{0}\pm|\boldsymbol{h}(\boldsymbol{k})|,\quad\Psi_{\pm}=\begin{pmatrix}\sqrt{1\pm\mbox{cos}\theta}\\
\pm e^{-i\phi}\sqrt{1\mp\mbox{cos}\theta}
\end{pmatrix}.
\end{equation}
It is evident that nonzero ${\cal J}_{a}$\,$-$\,${\cal J}_{b}$ and ${\cal D}_{z}$ can both open a gap at $\mbox{K}$ points, but the symmetries that they break are different. For ${\cal J}_{a}$\,$-$\,${\cal J}_{b}$, it breaks the inversion symmetry ${\cal P}$\,\cite{cheng2016spin, zyuzin2016magnon}, while ${\cal D}_{z}$ breaks the effective time reversal symmetry $C_{2x}{\cal T}$\,\cite{mook_PRX_2021_11, ni2025magnon}. Their different parity determines the magnon band topology, which can be further verified by the Berry curvature. Since two-dimensional\,(2D) cases only have $z$ component of Berry curvature, it can be expressed as
\begin{equation}\label{eq4}
\boldsymbol{{\Omega}}_{\pm}^{z}(\boldsymbol{k}) =\mp\frac{1}{2}\mbox{sin}\theta \left( \boldsymbol{\nabla}\theta \times \boldsymbol{\nabla}\phi \right).
\end{equation}
As shown in Fig.\,\ref{fig1}(c), the Berry curvature contributed by ${\cal D}_{z}$ is even with respect to $\boldsymbol{k}$, ${\Omega}(\boldsymbol{k})$\,$=$ \,${\Omega}(-\boldsymbol{k})$, leading to the topological nontrivial band with Chern number ${\cal C}$\,$=$\,sgn(${\cal D}_{z}$)\,\cite{bai2024coupled, bai2025dual,owerre2016first,zhu2021topological,mook_PRX_2021_11}. In contrast to ${\cal D}_{z}$, the Berry curvature contributed by  ${\cal J}_{a}$\,$-$\,${\cal J}_{b}$ shows odd parity, ${\Omega}(\boldsymbol{k})$\,$=$\,$-{\Omega}(-\boldsymbol{k})$, thus the magnon band is trivial with ${\cal C}$\,$=$\,$0$. This result is consistent with the symmetry analysis of the magnon Hamiltonian. The different parity of $\boldsymbol{\Omega}$ indicates the competition between ${\cal D}_{z}$ and ${\cal J}_{a}$\,$-$\,${\cal J}_{b}$, leading to the topological phase transition. According to Eq.\,(\ref{eq5}), the topological phase transition will occur when the gap\,($\Delta$) at the $\mbox{K}$ point closes, that is,
\begin{equation}\label{eq5}
    \frac{1}{2}\left({\cal J}_{a} - {\cal J}_{b}\right)f_{k} + {\cal D}_{z}d_{k} = 0 \big |_{k=\mbox{K},\mbox{K}^{\prime}},
\end{equation}
thereby, the phase boundary occurs at $\pm\frac{\sqrt{3}}{2}$\,$\left({\cal J}_{a}-{\cal J}_{b}\right)$\,$=$\,${\cal D}_{z}$. Given ${\cal J}_{a}$\,$-$\,${\cal J}_{b}$\,$=$\,$-6$\,${\cal E}_{z}$, the magnons of this bilayer can be switched between Chern and trivial insulating states by a finite ${\cal E}_{z}$ as shown in Fig.\,\ref{fig3}(b). Based on Eq.\,(\ref{eq5}) and Fig.\,\ref{fig3}(b), the critical ${\cal E}_{z}$ for bilayer $\mbox{CrSCl}$ and ${\mbox{CrSBr}}$ is $\pm{0.067}$\,$\mbox{eV/\AA}$ and $\pm{0.033}$\,$\mbox{eV/\AA}$, respectively.

Fig.\,\ref{fig3}(a) shows the magnon bands of bilayer $\mbox{CrSCl}$ under ${\cal E}_{z}$. Clearly, it exhibits non-equivariant energy dispersions at different valley-points, i.e., $\epsilon{(\mbox{K})}$\,$\neq$\,$\epsilon{(\mbox{K}^{\prime})}$. As discussed above, for the magnons in this bilayer, ${\cal E}_{z}$ breaks ${\cal P}$ and ${\cal D}_{z}$ breaks ${\cal T}^{\prime}$, jointly giving rise to valley polarization. This behavior is similar to the magnon valley polarization induced by the asymmetric spin exchanges in magnets with ${\cal P}$ symmetry breaking\,\cite{matsumoto2020nonreciprocal, wildes2021search, hayami2022essential, tong2016concepts}. Notably, at ${\cal E}_{z}$\,$=$\,$0.2$\,$\mbox{eV/\AA}$, the strength of valley polarization or nonreciprocity, defined as energy difference $\epsilon_{r}$\,$=$\, $\epsilon(\mbox{K})$\,$-$\,$\epsilon(-\mbox{K}^{\prime})$, reaches $3.5$\,$\mbox{meV}$, comparable to the energy scale of a $30$\,$\mbox{T}$ magnetic field. Importantly, since ${\cal D}_{z}$ remains invariant under reversal of ${\cal E}_{z}$, switching the sign of ${\cal J}_{a}$\,$-$\,${\cal J}_{b}$ via ${\cal E}_{z}$ reverses the sign of the valley polarization. This enables the electric field to fully dictate the sign and magnitude of valley polarization, in contrast to the case that relies on the SOC strength\,\cite{matsumoto2020nonreciprocal, wildes2021search, hayami2022essential}. 

In addition, the Berry curvature and orbital moments at each valley, as shown in Fig.\,\ref{fig3}(a), undergo a sign change upon reversal of the ${\cal E}_{z}$, highlighting the potential for electrically tunable magnon anomalous Hall effect in this bilayer magnons. Fig.\,S11\,\cite{Supplemental_Materials} shows the numerical result of thermal Hall conductivity $\alpha^{s}_{xy}$ and orbital Hall conductivity $\alpha^{o}_{xy}$ of bilayer $\mbox{CrSCl}$ as a function of ${\cal E}_{z}$. For fixed temperature, ${\cal E}_{z}$ can effectively modulate the thermal Hall coefficient, with the strength decreasing as ${\cal E}_{z}$ increases. Unlike $\alpha^{s}_{xy}$, $\alpha^{o}_{xy}$ is not monotonic in ${\cal E}_{z}$, initially decreasing and then increasing, characterized by a double-well curve. These results demonstrate the feasibility of achieving precise electric-field-controlled manipulation of magnon transport. 

\textit{Magnetic anisotropy dependence.} As discussed above, the magnetic anisotropy of bilayer $\mbox{CrSe}Y$\, ($Y$\,$=$\,$\mbox{Cl,\,Br})$ exhibits an easy-plane character, where the ${\cal T}^{\prime}$ symmetry (defined as $C_{2z}{\cal T}$\,\cite{mook_PRX_2021_11}) can be broken by in-plane DMI. Specifically, in bilayer $\mbox{CrSeBr}$, the magnetic anisotropy of $\mbox{Cr}^{3+}$ ions along $x$ axis is lower in energy than that along $y$ and $z$ axis by $0.005$ and $0.1$\,$\mbox{meV}$, respectively. Such magnetic anisotropy can stabilize the long range FM order along the $x$ axis\,\cite{mermin1966absence}, enabling its magnons to possess a ${\cal D}_{x}$ induced topological nontrivial gap at $\mbox{K}$ points. Due to the strong spin-layer coupling of bilayer $\mbox{CrSeBr}$, the difference between NN intralayer spin exchanges in adjacent layers under ${\cal E}_{z}$ is given by ${\cal J}_{a}$\,$-$\,${\cal J}_{b}$\,$=$\,$-7$\,${\cal E}_{z}$. The nonzero ${\cal J}_{a}$\,$-$\,${\cal J}_{b}$ breaks the ${\cal P}$ symmetry and opens a trivial gap of magnons, which competes with ${\cal D}_{x}$, driving the topological phase transition. According to Eq.\,(\ref{eq5}), the critical ${\cal E}_{z}$ for bilayer $\mbox{CrSeCl}$ is $\pm{0.08}$\,$\mbox{eV/\AA}$.

\begin{figure}
\centering
\includegraphics[scale=0.19]{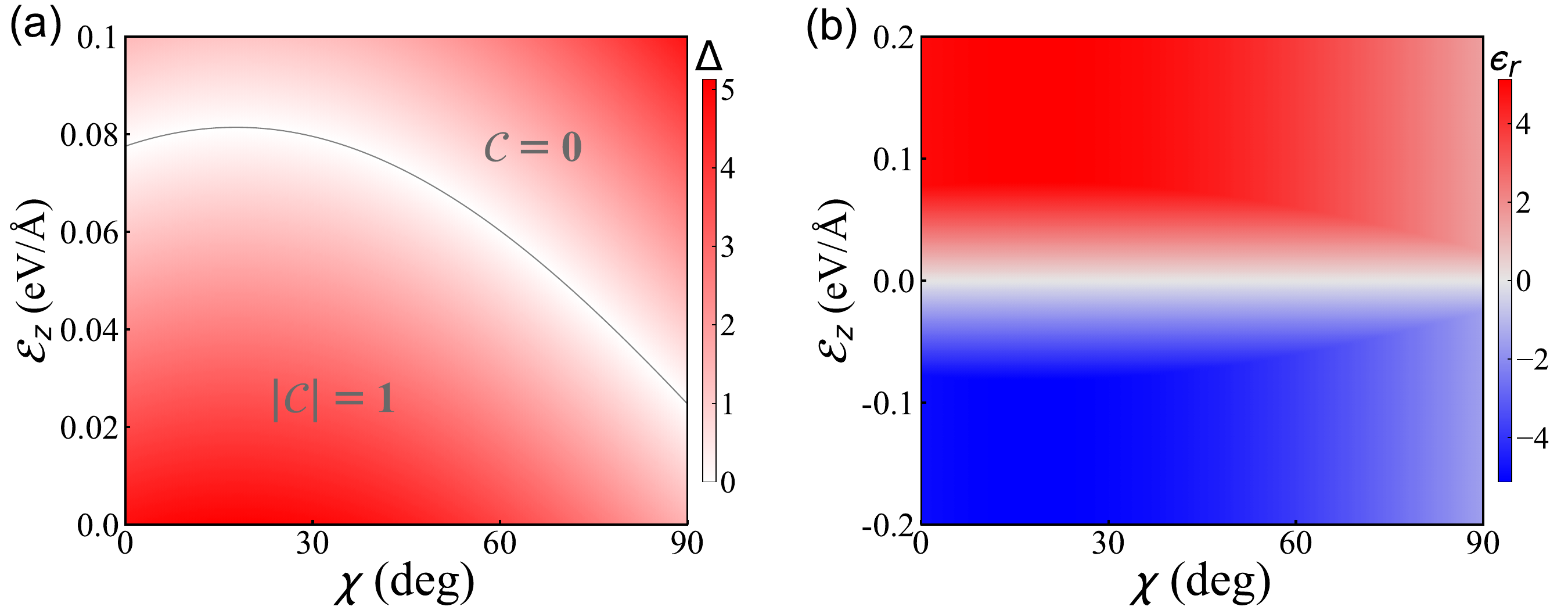}
    	\caption{The magnon phase diagram (a) and the valley polarization (nonreciprocity) distribution (b) in bilayer $\mbox{CrSeBr}$ as a function of ${\cal E}_{z}$ and $\chi$.}
\label{fig4}
\end{figure}

It's noteworthy that magnetic anisotropy between the $x$ axis and the $y$ axis is nearly degenerate, while the magnitudes of ${\cal D}_{x}$ and ${\cal D}_{y}$ exhibit distinct values, with ${\cal D}_{x}$\,$=$\,0.49\,$\mbox{meV}$ and ${\cal D}_{y}$\,$=$\,0.15\,$\mbox{meV}$, respectively. When the spin orientation lies in the $xy$-plane, the general formula for the phase boundary can be expressed as
\begin{equation}\label{eq6}
    {\cal D}_{x}\mbox{cos}\chi + {\cal D}_{y}\mbox{sin}\chi = \mp3.5\sqrt{3}\,{{\cal E}_{z}},
\end{equation}
where $\chi$ refers to the angle between the spin vector and the $x$ axis. The corresponding phase diagram is presented in Fig.\,\ref{fig4}(a), which clearly demonstrates a nonlinear and tunable phase boundary. In addition, since the spin orientation along the $y$ axis is only higher in energy than that along the $x$ axis by 0.005\,$\mbox{meV}$, which indicates that a weak magnetic field on the order of 10\,$\mbox{mT}$ is sufficient to effectively modulate the band topology. Meanwhile, the critical ${\cal E}_{z}$ decreases from $\pm{0.08}$\,$\mbox{eV/\AA}$ to $\pm{0.02}$\,$\mbox{eV/\AA}$ when spins are aligned along the $y$ axis. Similarly, a synergistic enhancement effect is also observed in the valley polarization of magnons. As shown in Fig.\,\ref{fig4}(b), under an applied electric field of 0.04\,$\mbox{eV/\AA}$, merely a 10\,$\mbox{mT}$ variation in the magnetic field can induce a 2\,$\mbox{meV}$ valley polarization. Compared with previous studies\,\cite{gitgeatpong2017nonreciprocal, iguchi2015nonreciprocal}, the synergistic coupling between magnetic and electric fields in this work significantly reduces the required external magnetic field strength for magnon manipulation by two orders of magnitude.  

In summary, we present a general strategy for electrical control of topological magnons in bilayer ferromagnetic insulators. Owing to the strong spin-layer coupling, a subtle and robust interplay emerges between the electric field and layer index, enabling precise manipulation of valley polarization, topological phases, and anomalous Hall transport of magnons. As exemplified in the bilayer $\mbox{Cr}XY$, the critical electric field is found to lie in the range of approximately $\pm0.05$$\sim$$\pm0.1$\,$\mbox{eV/\AA}$, well within the experimentally achievable window. More importantly, the synergistic coupling between magnetic and electric fields advances the manipulation of topological magnons, promoting spintronics with robust magnetoelectric coupling.

\textit{Acknowledgements.}
The authors thank Prof.\,Jian Zhou for helpful discussions. This work is supported by the National Natural Science Foundation of China (Grants No.\,12374092 and No.\,T2425029), Natural Science Basic Research Program of Shaanxi (Grants No.\,2023-JC-YB-017 and No.\,2022JC-DW5-02), Shaanxi Fundamental Science Research Project for Mathematics and Physics (Grant No. 22JSQ013), “Young Talent Support Plan” of Xi'an Jiaotong University, the Open Project of State Key Laboratory of Surface Physics (Grant No.\ KF2023\_06), and the Xiaomi Young Talents Program. L.\,B.\,thanks the Vannevar Bush Faculty Fellowship Grant No.\,N00014-20-1C2834 from the Department of Defense and Grant No.\,DMR-1906383 from the National Science Foundation Q-AMASE-i Program (MonArk NSF Quantum Foundry).

\textit{Data availability}. All data are available from the authors upon reasonable request.


\bibliography{main_ref}

\end{document}